\documentclass[journal]{IEEEtran}
\ifCLASSINFOpdf
\else
\fi
\usepackage{amsmath}
\usepackage{amsmath}
\usepackage{amssymb}
\usepackage{graphicx}
\usepackage{cite}
\usepackage{booktabs}
\usepackage{algorithm}
\usepackage{algorithmicx}
\usepackage{algpseudocode}
\usepackage{bm}
\usepackage{booktabs}
\usepackage{tabularx}
\usepackage{subcaption}
\usepackage{mathrsfs}
\usepackage{bm}
\usepackage{makecell}
\begin{document}
%
\title{ProMSC-MIS: Prompt-based Multimodal Semantic Communication for Multi-Spectral Image Segmentation}
%
%
%

\author{Haoshuo Zhang, Yufei Bo and 
        Meixia Tao,~\IEEEmembership{Fellow,~IEEE}
\thanks{Haoshuo Zhang, Yufei Bo and Meixia Tao are with the Department of Electronic Engineering and the Cooperative Medianet Innovation Center (CMIC), Shanghai Jiao Tong University, Shanghai 200240, China (e-mail: zhuiguang@sjtu.edu.cn; boyufei01@sjtu.edu.cn; mxtao@sjtu.edu.cn).

A preliminary version of this work was accepted at the IEEE/CIC ICCC, Shanghai, China, Aug. 2025 \cite{zhang2025prompt}.
}
}

\maketitle

\begin{abstract}
Multimodal semantic communication has great potential to enhance downstream task performance by integrating complementary information across modalities. This paper introduces ProMSC-MIS, a novel Prompt-based Multimodal Semantic Communication framework for Multi-Spectral Image Segmentation. It enables efficient task-oriented transmission of spatially aligned RGB and thermal images over band-limited channels. Our framework has two main design novelties. First, by leveraging prompt learning and contrastive learning, unimodal semantic encoders are pre-trained to learn diverse and complementary semantic representations by using features from one modality as prompts for another. Second, a semantic fusion module that combines cross-attention mechanism and squeeze-and-excitation (SE) networks is designed to effectively fuse cross-modal features. Experimental results demonstrate that ProMSC-MIS substantially outperforms conventional image transmission combined with state-of-the-art segmentation methods. Notably, it reduces the required channel bandwidth by 50\%--70\% at the same segmentation performance, while also decreasing the storage overhead and computational complexity by 26\% and 37\%, respectively. Ablation studies also validate the effectiveness of the proposed pre-training and semantic fusion strategies. Our scheme is highly suitable for applications such as autonomous driving and nighttime surveillance.

\end{abstract}

%

%
\IEEEpeerreviewmaketitle

\section{Introduction}

Empowered by advances in artificial intelligence (AI), semantic communication has emerged as a transformative paradigm to redefine how information is extracted and transmitted for intelligent communication services. Unlike traditional communication frameworks that prioritize bit-level accuracy, semantic communication focuses on conveying the essential meaning of raw data that is task-relevant. It thus can not only reduce channel bandwidth requirement, but also enhance downstream task performance for intelligence applications \cite{Zhangpingzongshu, Gündüz2023BeyondTransmittingBits, AIpowerd}. Recent studies have demonstrated the significant promises of semantic communication by leveraging different neural networks (NN) architectures in unimodal data transmission including text \cite{Xie2021DeepLearningEnabledSemanticCommunicationSystems, text-trans2}, speech \cite{SpeechRecong, speech_recog2}, image \cite{Denisimg,zhwimage,zhwimag2,wutongimg,imgniukai} and video \cite{jinshivideo, zhangpingvideo, deepwive}. In contrast, research on multimodal semantic communication remains in its nascent stages \cite{Crossmodal}.



While multimodal data provides richer semantic information and greater performance potential, it introduces new design challenges.
Recent studies on multimodal semantic communication have mainly focused on improving downstream task performance through better NN architecture design and cross-modal semantic fusion. For architecture design, the authors in \cite{U-DeepSC} introduce a unified  deep learning-based semantic communication system, named U-DeepSC, to serve different tasks with multiple modalities of data including text, speech and image.
Regarding semantic fusion, the work \cite{IEEEhowto:zhu2024multimodalfusionbasedmultitasksemantic} performs semantic fusion at the transmitter via segment embeddings.
In \cite{IEEEhowto:wang2024distributed}, semantic fusion is performed at the server side in a distributed semantic communication system for audio-visual parsing, where two terminals extract and transmit modality-specific features for downstream tasks. The work \cite{IEEEhowto:luo2024multimodal} further utilizes the channel as a medium for cross-modal semantic fusion and designs a multimodal multi-user semantic communication system for RGB and thermal images.
Besides architecture design and semantic fusion, emerging techniques such as generative AI and self-supervised learning are being adopted in multimodal semantic communication to improve performance and training efficiency \cite{IEEEhowto:chen2024generative}, \cite{Multi-Modal-Self-Supervised}.


\begin{table*}[h]
    \centering
    \caption{Description of Notations}
    \label{tab:notions}
    \renewcommand{\arraystretch}{1.05} 
    \normalsize
    \begin{tabularx}{\textwidth}{lX}
        \toprule
        \textbf{Notation} & \textbf{Description} \\
        \midrule
        $\bm{x^r}$, $\bm{x^t}, \bm{x^{r'}}, \bm{x^{t'}}$ & RGB images, thermal images and their preprocessed versions.\\
        $\bm{y^r_{RGB}}$, $\bm{y^t_{RGB}}$ & Output of RGB semantic encoder for $\bm{x^r}$ and $\bm{x^{t'}}$. \\
        $\bm{y^r_{THE}}$, $\bm{y^t_{THE}}$ & Output of thermal semantic encoder for $\bm{x^{r'}}$ and $\bm{x^t}$. \\
        $\bm{v_r}$, $\bm{v_t}$ & Output of RGB prompt projection and thermal prompt projection. \\
        $\bm{F_r^s}$, $\bm{F_t^s},\bm{F_r^c}$, $\bm{F_t^c}$ & Output of self-attention and cross-attention transformer block for RGB and thermal images. \\
        $\bm{M_r}$, $\bm{M_t}$ &  Cross-attention matrix of $\bm{F_r^c}$ and $\bm{F_t^c}$.\\
        $\bm{\alpha_r}$, $\bm{\alpha_t}$ &  Output of cross-attention module for RGB images and thermal images.   \\
        $\bm{z_s}$&  Semantic fusion feature.\\
        $\bm{b}$, $\bm{\hat{b}}$ & Transmitted bit sequence, received bit sequence.\\
        $\bm{p_s}$ & Probability mapping table from semantic fusion feature $\bm{z_s}$ to bit sequence $\bm{b}$. \\
        $L_v$, $L_s$, $L_b$ & Length values of the vector $\bm{v_r} / \bm{v_t}$, $\bm{z_s}$ and $\bm{b} / \bm{\hat{b}}$. \\
        $\bm{m}$, $\bm{\hat{m}}$ & Ground truth label, predicted label.\\
        $f_{RGB}(\cdot), f_{THE}(\cdot)$ & Mapping of RGB semantic encoder and thermal semantic encoder.\\
        $g_{RGB}(\cdot), g_{THE}(\cdot)$ & Mapping of RGB prompt projection and thermal prompt projection.\\
        $f_{SF}(\cdot), f_{B}(\cdot), f_{D}(\cdot)$ & Mapping of semantic fusion module, bit generator module and semantic decoder.\\
        $\bm{\theta_r,\theta_t,\phi_r,\phi_t,\varphi_s,\varphi_b,\psi}$ & Trainable parameters of $f_{RGB}(\cdot)$, $f_{THE}(\cdot)$, $g_{RGB}(\cdot), g_{THE}(\cdot)$, $f_{SF}(\cdot), f_{B}(\cdot)$ and $f_{D}(\cdot)$. \\
        \bottomrule
    \end{tabularx}
\end{table*}

Despite these advances, a critical gap persists in multimodal semantic communication: most existing studies focus on the semantic fusion strategies without analyzing each modality's contribution to task performance. 
Understanding the distinct role of each modality is crucial for designing effective unimodal encoders and fusion strategies.
For instance, modalities with higher contribution could be allocated more computational resources during encoding.
Moreover, even when the dominant modality is identified, the overall performance may still be limited  if other modalities provide redundant or conflicting information. 
As such, current multimodal semantic communication still faces two key problems: (i) How to quantify performance gains brought by each modality so as to guide the design of unimodal semantic encoders and cross-modal semantic fusion strategies. (ii) How to train the unimodal semantic encoders to learn complementary features while mitigating potential cross-modal interference.


To address the above problems, this paper introduces a novel pre-training strategy that combines prompt learning and contrastive learning, where one modality serves as prompts to guide semantic feature extraction of another. Prompt learning is widely applied in the field of natural language processing. Its core idea is to reframe downstream tasks as generation or cloze questions, which are well-suited for pre-training models such as BERT \cite{bert} and GPT \cite{gpt}. We adopt this concept by transforming data from one modality into a ``prompt'', which provides cues for the feature extractor of the other modality. This prompt-guided process is optimized using contrastive learning, which encourages the unimodal encoders to produce more diverse features by maximizing their distance in the embedding space.

Building on the above foundation, we propose ProMSC-MIS, a prompt-based multimodal semantic communication system for multi-spectral image segmentation. Multi-spectral image segmentation plays a critical role in applications such as autonomous driving and nighttime surveillance, where images from the same viewpoint are captured by different sensors across different spectral ranges, such as RGB and thermal. While conventional segmentation methods based on convolutional neural networks (CNNs) \cite{IEEEhowto:ha2017mfnet, IEEEhowto:sun2019rtfnet, IEEEhowto:deng2021feanet} operate directly on source images without considering the communication constraints, ProMSC-MIS exploits semantic communication principles to take into account source transmission over practical band-limited wireless channels. Specifically, our framework first employs two pre-trained unimodal semantic encoders based on prompt learning and contrastive learning as mentioned above for semantic feature extraction from the RGB images and the thermal images, respectively. These features are then fused via a semantic fusion module, which enables the system to exploit the complementary strengths of each modality and enhance task performance. The fused representation is then transmitted over the band-limited and noisy wireless channel. Upon receiving the signals at the channel output, our framework then conducts image segmentation by using a semantic decoder that can be trained in an end-to-end manner.

\begin{figure*}[t]
  \centering
  \includegraphics[width=\linewidth]{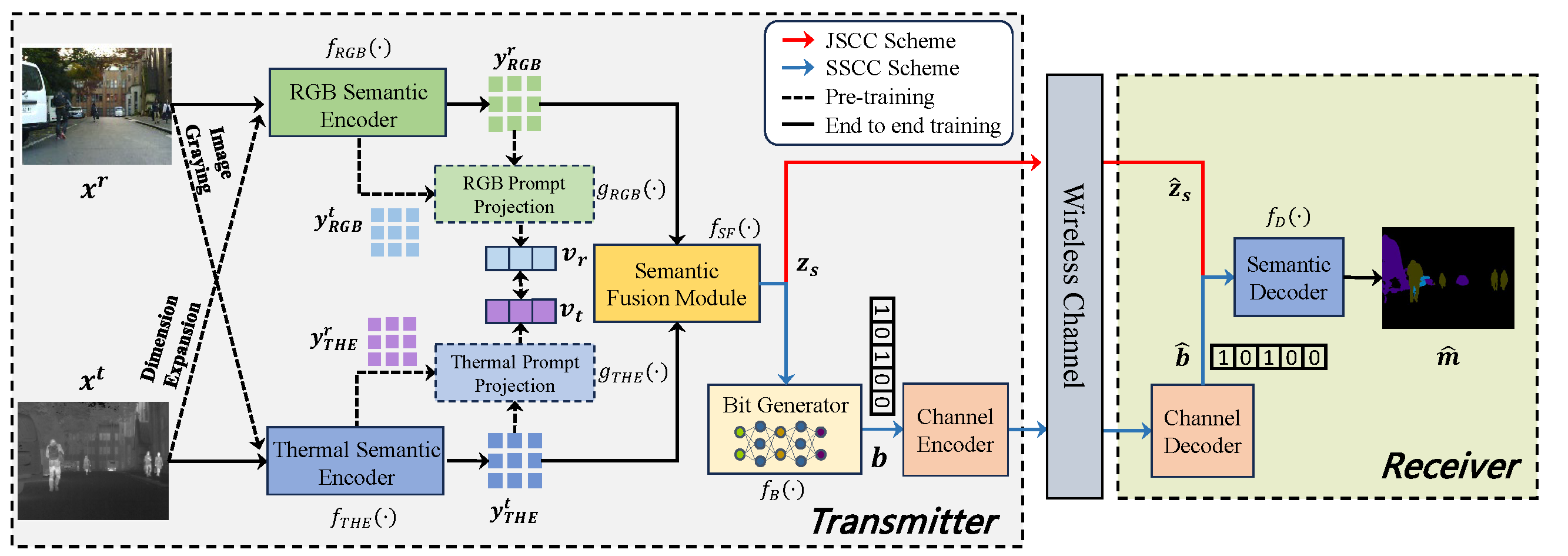}
  \caption{Framework of the proposed ProMSC-MIS. 
  }
  \label{fig:systemmodel}
\end{figure*}

The contributions of this paper are summarized as follows:

\begin{itemize}
    \item \textbf{ProMSC-MIS framework:} We propose ProMSC-MIS, a prompt-based multimodal semantic communication framework for multi-spectral image segmentation. By leveraging spatially aligned RGB and thermal image pairs, the system enables efficient and task-driven transmission over band-limited  channels.
    
    \item \textbf{Pre-training and fusion strategies:} We introduce a novel pre-training algorithm for learning diverse semantic features and further design a semantic fusion module to fuse them. Based on prompt learning and contrastive learning, the pre-training algorithm distinctively utilizes images from other modalities as prompts to guide unimodal encoders in capturing richer features. Subsequently, the semantic fusion module leverages a cross-attention mechanism and a self-attention-based network to enrich unimodal representations, enabling robust and effective semantic fusion.
     

    \item \textbf{Experimental validation:} 
    Experimental results demonstrate the efficiency and effectiveness of ProMSC-MIS. Compared with the conventional image transmission combined with state-of-the-art segmentation methods, ProMSC-MIS reduces the required channel bandwidth by 50\%--70\% at the same segmentation performance, while also decreasing the storage overhead and computational complexity by 26\% and 37\%, respectively. Ablation studies also validate the effectiveness of the proposed pre-training and semantic fusion strategies in the ProMSC-MIS framework.  Additionally, we analyze modality contributions across different bandwidth, offering insights for unimodal semantic encoders and fusion strategy design.
    
\end{itemize}

The remainder of this paper is organized as follows: Section II outlines the overall framework of ProMSC-MIS. Section III details its key module designs and the training strategy. Section IV presents and discusses the experimental results. Finally, Section V concludes the paper.

Notations: Throughout this paper, we denote the set of  $ m \times n $ real matrices as $\mathbb{R}^{m\times n}$. Scalars are represented  by lower case letters (e.g., $x$), vectors by boldface lower-case letters boldface case letters (e.g., $\bm{v}$), and matrices by upper-case letters (e.g., $\bm{M}$). The $l_2$-norm of a vector $\bm{v}$ is denoted by $\lVert \bm{v} \lVert_2$. The symbol $\odot$ represents the element-wise product. Key notations used frequently in this paper are summarized in Table \ref{tab:notions}.

\section{Overall Framework of ProMSC-MIS}

Fig. \ref{fig:systemmodel} shows the overall framework of ProMSC-MIS, where the transmitter has a pair of RGB and thermal images collected from different sensors to send to the receiver over a wireless channel for image segmentation. 

Let us denote the input RGB image and thermal image as $\bm{x^r} \in \mathbb{R}^{H \times W \times 3}$ and $\bm{x^t} \in \mathbb{R}^{H \times W \times 1}$, respectively, where $H$ and $W$ represent the height and the width of the image. At the transmitter, the input images are processed separately by two corresponding  unimodal semantic encoders. The RGB semantic encoder $f_{RGB}(\cdot)$ maps $\bm{x^r}$ to the RGB semantic features $\bm{y^r_{RGB}}\in \mathbb{R}^{H_r\times W_r \times C_r}$ while the thermal semantic encoder $f_{THE}(\cdot)$ maps $\bm{x^t}$ to the thermal semantic features $\bm{y^t_{THE}}\in \mathbb{R}^{H_t \times W_t \times C_t}$.  Here the two unimodal semantic encoders are pre-trained jointly by the proposed pre-training approach, with the assistance of two prompt projection modules (in dashed lines in Fig.~\ref{fig:systemmodel}). This is one of the main design novelties of this paper and shall be explained in detail in Section III-D. The extracted unimodal features, $\bm{y^r_{RGB}}$ and $\bm{y^t_{THE}}$, are then fed into a semantic fusion module $f_{SF}(\cdot)$ to generate a fused semantic representation $\bm{z_s}\in \mathbb{R}^{L_s}$ for channel transmission, where $L_s$ denotes the length of the semantic fusion vector. 

For channel transmission, ProMSC-MIS considers two different methods. One is joint source-channel coding (JSCC) (in red lines in Fig.~\ref{fig:systemmodel}), where the real-valued semantic representation $\bm{z_s}$ is transmitted directly after power normalization over $L_s$ channel uses and received as $\bm{\hat{z}_s}$. The other is separate source-channel coding (SSCC) (in blue lines in Fig.~\ref{fig:systemmodel}), where $\bm{z_s}$ is first digitized into a bit sequence $\bm{b} \in \mathbb{R}^{L_b}$ via a bit generator $f_B(\cdot)$, then sent to the channel after channel encoding and received as $\bm{\hat{b}}$ after channel decoding at the receiver side. Here $L_b$ denotes the length of the bit sequence $\bm{b}$.  

At the receiver, a semantic decoder $f_{D}(\cdot)$ performs image segmentation task using the received $\bm{\hat{z}_s}$ for the JSCC design, or the recovered bit sequences $\bm{\hat{b}}$ for the SSCC design. The final segmentation result is denoted as $\bm{\hat{m}} \in \mathbb{R}^{H \times W \times N}$, where $N$ represents the number of segmentation classes. The corresponding ground truth is denoted by $\bm{m}$. 

As stated above, our proposed ProMSC-MIS works for both JSCC and SSCC designs. To ensure better compatibility with existing communication systems, we adopt the SSCC design for the rest of the paper. In this case, the combined effects of channel coding, wireless transmission and channel decoding are modeled  as a Binary Symmetric Channel (BSC) \cite{bao2024sdacsemanticdigital}, where each transmitted bit $b_i$ is flipped with a certain transition probability, denoted as $p$. 

\section{Key Module Design and Training Strategy}
This section outlines key module design of ProMSC-MIS, including semantic encoder/decoder, semantic fusion module and bit generator in Section III-A, Section III-B and Section III-C, respectively. Then Section III-D details our training strategy.

\subsection{Semantic Encoder and Semantic Decoder}

\begin{figure}[t] 
    \centering
    \begin{minipage}{0.495\textwidth}
        \centering
        \includegraphics[width=\linewidth]{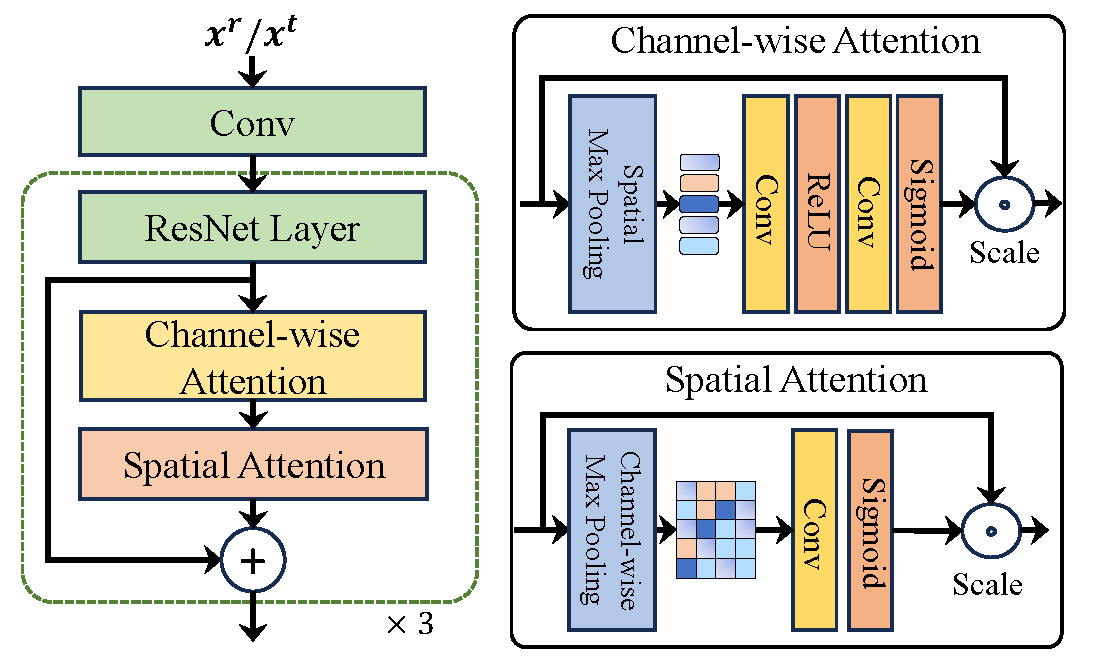} 
        \subcaption{The architecture of the semantic encoder.
 }\label{fig:encoder}
    \end{minipage}
    \begin{minipage}{0.4\textwidth}
        \centering
        \includegraphics[width=\linewidth]{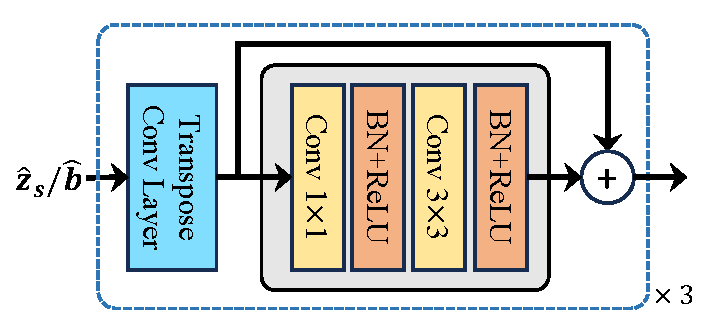} 
        \subcaption{The architecture of the semantic decoder.}\label{fig:decoder}
    \end{minipage}
    \caption{The architecture of the semantic encoder/decoder.}
    \label{fig:semantic_encoder}
\end{figure}

We employ the same network architecture for both the RGB semantic encoder $f_{RGB}(\cdot)$ and the thermal semantic encoder $f_{THE}(\cdot)$, as shown in Fig.~\ref{fig:semantic_encoder}(a).  It adopts ResNet-152 as backbone, and uses channel-wise attention and spatial attention modules to enhance feature extraction, as in \cite{IEEEhowto:deng2021feanet}. For channel-wise attention, spatial max pooling transforms input features into a channel descriptor vector. This vector is then processed by convolution layers employing ReLU and sigmoid activation functions to generate channel-specific weights, which then scale the input feature map to recalibrate channel importance. For spatial attention, channel-wise max pooling converts input features into a 2D spatial map highlighting prominent features. This map is then fed through convolution layers and a sigmoid function to produce a spatial attention map, which subsequently multiplies the input features element-wise to emphasize significant spatial regions.
Unlike \cite{IEEEhowto:deng2021feanet}, we remove the interaction between feature maps in this unimodal feature extraction stage, enabling the system to perform unimodal semantic feature extraction in the absence of data from the other.

Fig.~\ref{fig:semantic_encoder}(b) shows the architecture of the semantic decoder $f_{D}(\cdot)$. It is composed of several transpose convolution layers to progressively upsample the feature maps, ultimately mapping the semantic features to the corresponding segmentation results  $\bm{\hat{m}}$. Depending on whether JSCC or SSCC  is adopted, $f_{D}(\cdot)$ receives either the analog $\bm{\hat{z}_s}$ or the bit sequence $\bm{\hat{b}}$.

\subsection{Semantic Fusion Module}
\begin{figure}[t] 
    \centering
    \begin{minipage}{0.495\textwidth}
        \centering
        \includegraphics[width=\linewidth]{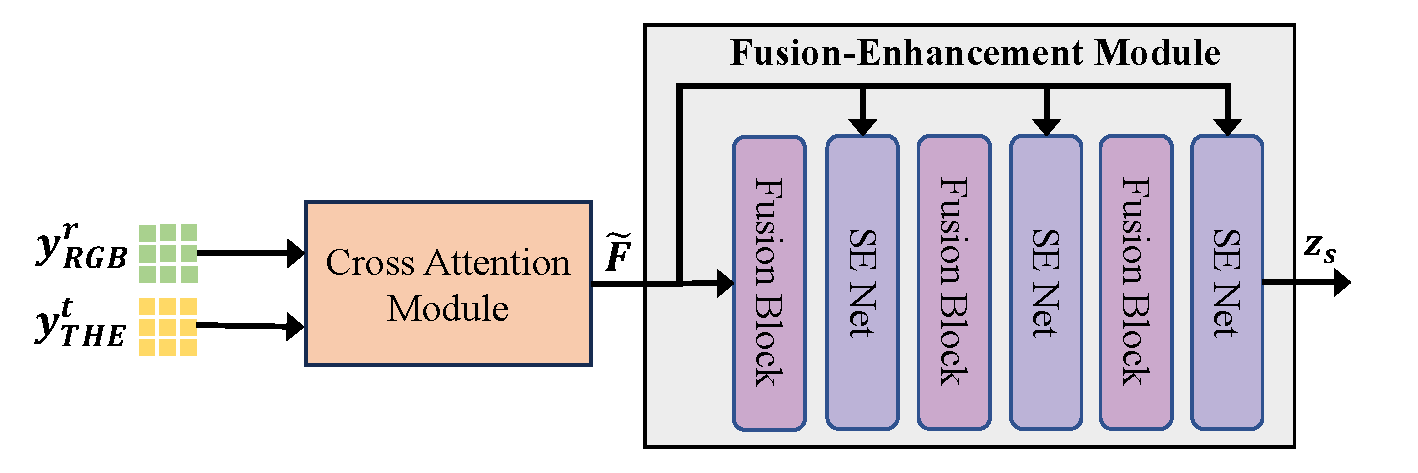} 
        \subcaption{The architecture of the semantic fusion module.
 }\label{fig:fusion_sub1}
    \end{minipage}
    \begin{minipage}{0.495\textwidth}
        \centering
        \includegraphics[width=\linewidth]{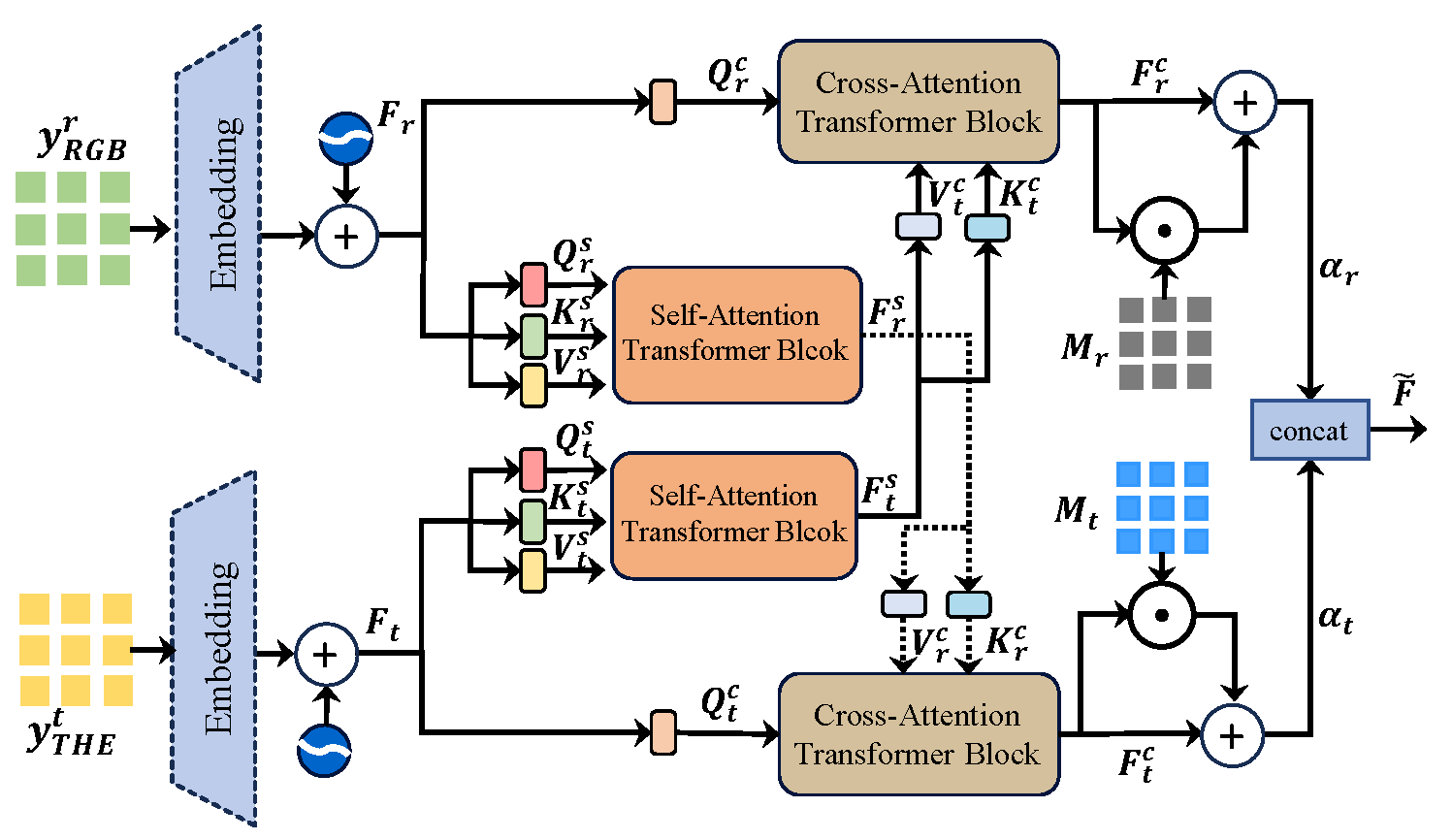} 
        \subcaption{The architecture of the cross-attention module.}\label{fig:fusion_sub2}
    \end{minipage}
    \begin{minipage}{0.495\textwidth}
        \centering
        \includegraphics[width=\linewidth]{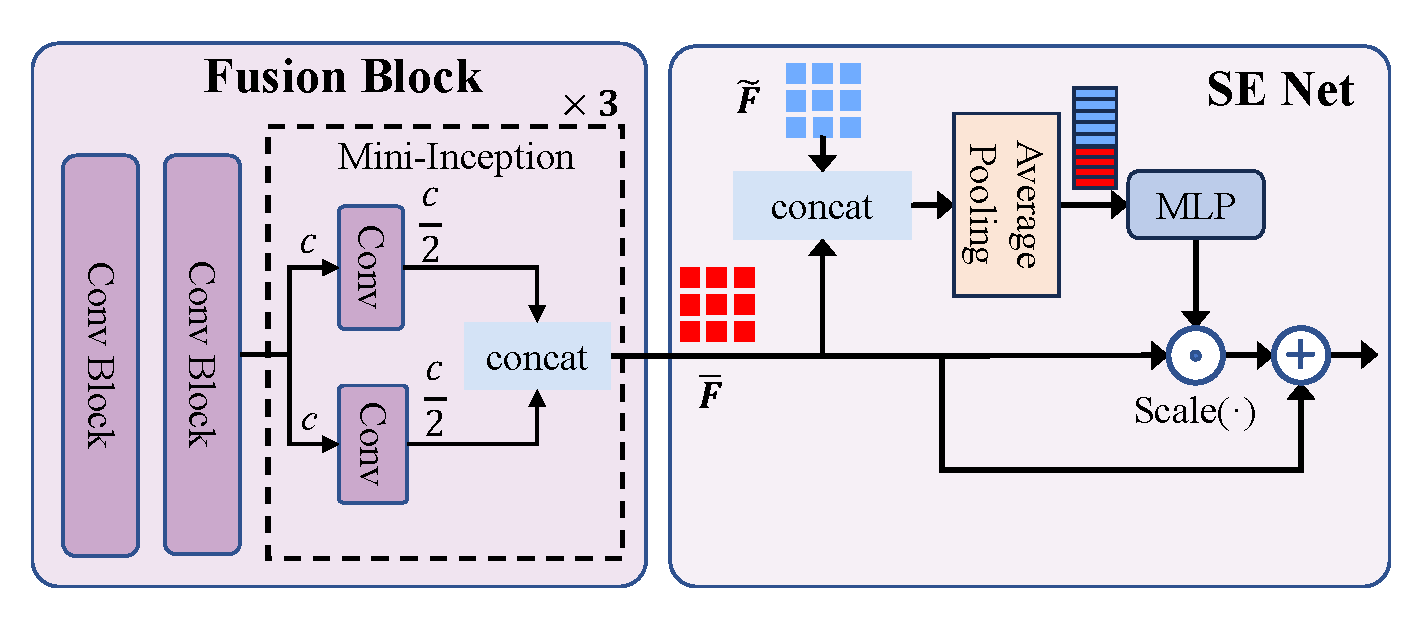} 
        \subcaption{The architecture of the fusion-enhancement module.}\label{fig:fusion_sub3}
    \end{minipage}
    \caption{The architecture and design details of the semantic fusion module.}
    \label{fig:fusion}
\end{figure}

After unimodal semantic encoding, a semantic fusion module is designed to effectively fuse the extracted unimodal features. 
Its structure is shown in Fig.~\ref{fig:fusion}(a), composed of a cross-attention module and a fusion-enhancement module. The cross-attention module first takes unimodal features $\bm{y_{RGB}^r}$ and $\bm{y_{THE}^t}$ as input to produce $\bm{\tilde{F}}$, which is then fed into the fusion-enhancement module to obtain the final fused feature $\bm{z_s}$.

Specifically, the cross-attention module comprises several transformer blocks \cite{vaswani2023attentionneed}, as illustrated in Fig.~\ref{fig:fusion}(b). The inputs $\bm{y_{RGB}^{r}}$ and $\bm{y_{THE}^{t}}$ are first processed through embedding and positional encoding to generate $\bm{F_{r}}\in \mathbb{R}^{L_r\times D}$ and $\bm{F_{t}}\in \mathbb{R}^{L_t\times D}$, respectively. These representations are then fed into a self-attention transformer block followed by a cross-attention transformer block, as detailed below.

The self-attention transformer block maps the input feature $\bm{F_r}$ to an output feature $\bm{F_r^s}$ using two main sub-layers: a multi-head self-attention (MHSA) mechanism and a feed-forward network (FFN). Each sub-layer incorporates a residual connection followed by layer normalization, structured as 
\begin{equation}
\begin{aligned}
\bm{F_r'} = \text{LayerNorm}(\bm{F_r} + \text{MHSA}(\bm{F_r})),
\end{aligned}
\end{equation}
\begin{equation}
\begin{aligned}
\bm{F_r^s} =  \text{LayerNorm}(\bm{F_r'}+\text{FFN}(\bm{F_r'})).
\end{aligned}
\end{equation}
For the MHSA sub-layer, the input feature $\bm{F_r}$ is first linearly transformed to produce the query $\bm{Q_r^s}$, key $\bm{K_r^s}$, and value $\bm{V_r^s}$ matrices, all of dimension $\mathbb{R}^{L_r\times D_m}$, i.e.,
\begin{equation}
\bm{Q_r^s}=\bm{F_r}\bm{W_{Q,r}^s}, \quad \bm{K_r^s}=\bm{F_r}\bm{W_{K,r}^s}, \quad \bm{V_r^s}=\bm{F_r}\bm{W_{V,r}^s}, \label{eq:rev_qkv_projection}
\end{equation}
where $\bm{W_{Q,r}^s}$, $\bm{W_{K,r}^s}$ and $\bm{W_{V,r}^s} \in \mathbb{R}^{D \times D_m}$ are learnable weight matrices. The MHSA sub-layer comprises $H$ parallel self-attention (SA) heads. The outputs of these heads are concatenated and subsequently projected by a linear transformation matrix $\bm{W_r^s}\in \mathbb{R}^{D_m\times D}$, defined as 
\begin{equation}
\begin{aligned}
\text{SA}_{h}(\bm{F_r}) = \text{softmax}(\frac{\bm{Q_{r,h}^s} \bm{K_{r,h}^{sT}}}{\sqrt{D_m / H}})\bm{V_{r,h}^{s}},
\end{aligned}
\end{equation}
\begin{equation}
\begin{aligned}
\text{MHSA}(\bm{F_r}) = [\text{concat}_{h=1}^H\text{SA}_{h}(\bm{F_r})]\bm{W^{s}_r}.
\label{eq:rev_mhsa_output}
\end{aligned}
\end{equation}
where $\bm{Q_{r,h}^s}$, $\bm{K_{r,h}^s}$ and $\bm{V_{r,h}^s} \in \mathbb{R}^{L_r \times \frac{D_m}{H}}$ are derived from $\bm{Q_r^s}$, $\bm{K_r^s}$ and $\bm{V_r^s}$, respectively, through dedicated linear transformations for each head. Through the self-attention transformer block, the RGB modality feature $\bm{F_r}$ is transformed into $\bm{F_r^s}$. Similarly, the thermal modality feature $\bm{F_t}$ is processed through an identical transformer block structure, yielding $\bm{F_t^s}$. 

Following the self-attention transformer block, a cross-attention transformer block is employed to fuse cross-modal information between the modalities. Specifically, $\bm{F_r^s}$ is linearly projected to generate the query matrix $\bm{Q_r^c}\in \mathbb{R}^{L_r\times D_m}$. Concurrently, the thermal features $\bm{F_t^s}$ are transformed into the key matrix $\bm{K_t^c}\in \mathbb{R}^{L_t \times D_m}$ and the value matrix $\bm{V_t^c}\in \mathbb{R}^{L_t \times D_m}$. These matrices serve as inputs to the cross-attention transformer block. Its structure is similar to the self-attention blocks, with the self-attention component replaced by multi-head cross-attention (MHCA). The updated RGB feature $\bm{F_r^c}$ is obtained as follows: 
\begin{equation}
\begin{aligned}
\bm{{F_r^s}'}= \text{LayerNorm}(\bm{F_r^s} + \text{MHCA}(\bm{F_r^s})),
\end{aligned}
\end{equation}
\begin{equation}
\begin{aligned}
\bm{F_r^c} = \text{LayerNorm}(\bm{{F_r^s}'}+\text{FFN}(\bm{{F_r^s}'})),
\end{aligned}
\end{equation}
Here, the $\text{MHCA}(\cdot)$ operation aggregates information from $\bm{V_t^c}$ based on the similarity $\bm{Q_r^c}$ and $\bm{K_t^c}$. The MHCA mechanism itself is defined as
\begin{equation}
\begin{aligned}
\text{CA}_{h}(\bm{F_r^s}) = \text{softmax}(\frac{\bm{Q_{r,h}^c} \bm{K_{t,h}^{cT}}}{\sqrt{D_m / H}})\bm{V_{t,h}^{c}},
\end{aligned}
\end{equation}
\begin{equation}
\begin{aligned}
\text{MHCA}(\bm{F_r^s}) = [\text{concat}_{h=1}^H\text{CA}_{h}(\bm{F_r^s})]\bm{W^c_r},
\end{aligned}
\end{equation}
where $\bm{W_r^c}\in \mathbb{R}^{D_m\times D}$ is an output linear transformation matrix. The head-specific matrices $\bm{Q_{r,h}^c}$, $\bm{K_{t,h}^c}$, and $\bm{V_{t,h}^c}\in \mathbb{R}^{L_t\times\frac{D_m}{H}}$, are obtained through dedicated linear transformations of $\bm{Q_{r}^c}$, $\bm{K_{t}^c}$, and $\bm{V_{t}^c}$, respectively, for each of the $H$ attention heads. The output $\bm{F_r^c}\in \mathbb{R}^{L_t\times D}$ represents the cross-attention enhanced RGB features.

Furthermore, to mitigate potential negative impact from the other modality, we introduce a learnable matrix $\bm{M_r}\in \mathbb{R}^{L_r\times D}$. It can dynamically adjust the contribution of the cross-modal information, refining $\bm{F_r^c}$ into the final RGB representation $\bm{\alpha_r}$, i.e.,
\begin{equation}
\bm{\alpha_r} = \bm{F_r^c} \odot \bm{M_r} + \bm{F_r^c}. \label{eq:cross_gating}
\end{equation}
A similar process is applied to the thermal features $\bm{F_t^s}$ (using $\bm{F_r^s}$ to generate key and value matrices)  to obtain the corresponding enhanced thermal output $\bm{\alpha_t}\in\mathbb{R}^{L_t\times D}$. The resulting $\bm{\alpha_r}$ and $\bm{\alpha_t}$ are then concatenated to form $\bm{\widetilde{F}}$, which serves as the output of the cross-attention module and is then fed into the fusion-enhancement module.

The fusion-enhancement module is composed of an alternating sequence of fusion blocks and squeeze-and-excitation (SE) networks \cite{SENet}, with their architectures detailed in Fig.~\ref{fig:fusion}(c). Each fusion block consists of convolutional blocks and mini-inception modules. These mini-inception modules operate by splitting the input feature map into two parts along the channel dimension. This approach is designed to enhance feature diversity, enabling the model to capture multi-scale information more effectively and improve the robustness of feature fusion \cite{IEEEhowto:ha2017mfnet}. The output of each fusion block, denoted as $\bm{\bar{F}}$, is then processed by the SE network, where $\bm{\bar{F}}$ and $\bm{\widetilde{F}}$ are concatenated and passed through a global average pooling layer to aggregate contextual information.  The pooled representations are fed into a multilayer perceptron (MLP) to generate channel-wise attention weights, which are subsequently applied to $\bm{\bar{F}}$ via element-wise multiplication. After three successive stages of fusion blocks combined with the SE networks above, the final semantic fusion feature $\bm{z_s}$ is obtained.

\subsection{Bit Generator}
The mapping from the fused semantic feature $\bm{z_s}$ to the bit sequence $\bm{b}$ is a key component to ensure compatibility with existing communication systems. Intuitively, this can be done by an element-wise uniform quantizer, which however may suffer performance loss. In this paper, we introduce a learnable bit generator to implement such mapping.

Specifically, the fused semantic feature $\bm{z_s}$ is fed into a probabilistic generative layer to produce a probability table $\bm{p_s} \in \mathbb{R}^{L_b \times 2}$. The $\ell$-th row of $\bm{p_s}$, for $\ell$ =1, \dots, $L_b$, represents the probability of being 0 and 1 for the $\ell$-th bit of the sequence.
To obtain a discrete bit sequence $\bm{b}$ from these probabilities, a sampling step is necessary. A straightforward approach is to make hard decisions according to the probabilities. However, this method is non-differentiable, making it difficult to train the neural network.  To address this problem, we adopt the Gumbel-Softmax trick for probabilistic sampling as did in \cite{Bo-digital}. 

Before concluding this subsection, we would like to remark that the proposed learnable bit generator that maps the fused semantic feature $\bm{z_s}$ of dimension $L_s$ to the  bit sequence $\bm{b}$ of the length $L_b$ is not a simple $L_b/L_s$-bit quantizer, but a differential binary sampling process. If not mentioned otherwise, we assume $L_s = L_b$ throughout the experimental simulation. 


\subsection{Training Strategy}
The training of ProMSC-MIS consists of two stages, pre-training of the unimodal semantic encoders, followed by end-to-end training of the entire system. Algorithm \ref{alg:training} demonstrates the specific training process.

\subsubsection{Pre-training of unimodal semantic encoders} To enhance the unimodal semantic encoders' ability to extract richer semantic features under a limited number of parameters, we propose a pre-training method based on prompt learning and contrastive learning. The core idea is to utilize images from one modality as  prompts for the encoder of the other modality. This encourages the encoders to focus on capturing features that are complementary across modalities instead of redundant common information.

Firstly, the unimodal semantic encoders process their corresponding inputs: the RGB semantic encoder $f_{RGB}(\cdot)$ maps an RGB image $\bm{x^r}$ to its feature representation $\bm{y_{RGB}^r}$, and similarly, the thermal semantic encoder $f_{THE}(\cdot)$ maps a thermal image $\bm{x^t}$ to $\bm{y_{THE}^t}$. Before cross-modal prompting, input images undergo a preprocessing step since they have different color channel depth. That is, RGB images $\bm{x^r}$ are converted to grayscale $\bm{x^{r'}}\in \mathbb{R}^{H\times W \times 1}$, while thermal images $\bm{x^t}$ are expanded to three channels, yielding $\bm{x^{t'}}\in \mathbb{R}^{H\times W \times 3}$. These processed images then serve as prompts to the other modality's encoder, i.e., 

\begin{equation}
\begin{aligned}
\bm{y^t_{RGB}} &= f_{RGB}(\bm{{x^{t}}'};\bm{\theta_r}) \in \mathbb{R}^{H_r \times W_r \times C_r},
\end{aligned}
\end{equation}
\begin{equation}
\begin{aligned}
\bm{y^r_{THE}} &= f_{THE}(\bm{{x^{r}}'};\bm{\theta_t}) \in \mathbb{R}^{H_t \times W_t \times C_t},
\end{aligned}
\end{equation}
where $\bm{\theta_r}$ and $\bm{\theta_t}$ represent the network parameters of $f_{RGB}(\cdot)$ and $f_{THE}(\cdot)$, respectively.

Next, to align the output features of the two semantic encoders, a prompt projection operation is introduced. We define the RGB prompt projection as $g_{RGB}(\cdot)$ and the thermal prompt projection as $g_{THE}(\cdot)$, parameterized by $\bm{\phi_r}$ and $\bm{\phi_t}$, respectively. Each projection module consists of several convolutional layers followed by an MLP. 
For each modality, we concatenate the features extracted from its original input and the features extracted when prompted by the other modality. The concatenated features are projected into a shared semantic space by $g_{RGB}(\cdot)$ and $g_{THE}(\cdot)$, i.e., 
\begin{equation}
\begin{aligned}
\bm{v_r} &= g_{RGB}(\mathrm{concat}(\bm{y^r_{RGB}},\bm{y^t_{RGB}});\bm{\phi_r})\in \mathbb{R}^{L_v},
\end{aligned}
\end{equation}
\begin{equation}
\begin{aligned}
\bm{v_t} &= g_{THE}(\mathrm{concat}(\bm{y^t_{THE}},\bm{y^r_{THE}});\bm{\phi_t})\in \mathbb{R}^{L_v},
\end{aligned}
\end{equation}
where $\bm{v_r}$ and $\bm{v_t}$ are the projection vectors in $\mathbb{R}^{L_v}$. 
The objective is to make these projected vectors as dissimilar as possible, thereby encouraging the encoders to capture complementary aspects. This is achieved by minimizing their cosine similarity, given by:
\begin{equation}
\begin{aligned}
     \mathcal{L}_v(\bm{v_r},\bm{v_t}) &= \frac{|\bm{v_r} \cdot \bm{v_t}|}{\lVert \bm{v_r} \lVert_2 \cdot \lVert \bm{v_t} \lVert_2} \in (0,1),
\end{aligned}
\end{equation}
where $\lVert \cdot \rVert_2$ represents the $l_2$-norm.

This pre-training strategy equips each unimodal encoder with prior knowledge of the other modality, thereby enhancing its ability to extract complementary features under a limited number of parameters.

\begin{algorithm}[t]
\caption{Two-phase training algorithm}
\label{alg:training}
\begin{algorithmic}[1]
    \State \textbf{Input:} Training datasets $\bm{X}$ consisting of RGB-T image pairs. Image segmentation labels. The number of training epochs for two phases, $N_1$ and $N_2$, the weight factor $\lambda$, the learning rate.
    \State \textbf{Output:} Optimized the neural network parameters $\{\bm{\theta_r^*},\bm{\theta_t^*},\bm{\phi_r^*},\bm{\phi_t^*},\bm{\varphi_s^*},\bm{\varphi_b^*},\bm{\psi^*} \}$ 
    \State \textbf{First Phase:}
    \For{$i \leftarrow 1$ \textbf{to} $N_1$}
    \State Choose a batch of samples in $\bm{X}$.
    \State Obtain RGB images $\bm{x^r}$ and thermal images $\bm{x^t}$.
    \State Execute $\bm{y_{RGB}^{r}}=f_{RGB}(\bm{x^r})$, $\bm{y_{THE}^{t}}=f_{THE}(\bm{x^t})$.
    \State Transform dimensions and obtain $\bm{{x^{r}}'}$ and $\bm{{x^{t}}'}$.
    \State Execute $\bm{y_{RGB}^{t}}=f_{RGB}(\bm{{x^{t}}'})$, $ \bm{y_{THE}^{r}}=f_{THE}(\bm{{x^{r}}'})$.
    \State Execute $\bm{v_r} = g_{RGB}(\mathrm{concat}(\bm{y^r_{RGB}},\bm{y^t_{RGB}}))$, $\bm{v_t}=g_{THE}(\mathrm{concat}(\bm{y^t_{THE}},\bm{y^r_{THE}}))$.
    \State Compute pre-training loss $\mathcal{L}_v(\bm{v_r,v_t})$.
    \State Update parameters $\{\bm{\theta_r},\bm{\theta_t},\bm{\phi_r},\bm{\phi_t}\}$.
    \EndFor
    \State \textbf{Second Phase:}
    \State Load the parameters 
    $\{\bm{\theta_r},\bm{\theta_t}\}$ trained in the first phase.
    \For{$i \leftarrow 1$ \textbf{to} $N_2$}
    \State Choose a batch of samples in $\bm{X}$.
    \State Obtain RGB images $\bm{x^r}$ and thermal images $\bm{x^t}$.
    \State Execute  $\bm{y_{RGB}^{r}}=f_{RGB}(\bm{x^r})$, $ \bm{y_{THE}^{t}}=f_{THE}(\bm{x^t})$.
    \State Execute semantic fusion $\bm{z_s} = f_{SF}(\bm{y_{RGB}^{r}},\bm{y_{THE}^{t}})$.
    \State Generate discrete bits $\bm{b}=f_B(\bm{z_s})$.
    \State Transmit $\bm{b}$ through BSC and received as $\bm{\hat{b}}$.
    \State Execute semantic decoding $\bm{\hat{m}}=f_D(\bm{\hat{b}})$.
    \State Compute end-to-end training loss $\mathcal{L}_e(\bm{\hat{m}},\bm{m})$.
    \State Update parameters 
    $\{\bm{\theta_r},\bm{\theta_t},\bm{\varphi_s},\bm{\varphi_b},\bm{\psi}\}$
    \EndFor
\end{algorithmic}
\end{algorithm}

\subsubsection{End-to-end training of the entire ProMSC-MIS} 
After pre-training, the whole network is then optimized in an end-to-end manner. The optimization is guided by a composite loss function $\mathcal{L}_e$, which linearly combines the Diceloss $\mathcal{L}_{Dice}$ \cite{Milletari2016VNet} and a Soft Cross-Entropy loss $\mathcal{L}_{SoftCE}$, i.e.,
\begin{equation}
\begin{aligned}
\mathcal{L}_e(\bm{\hat{m}},\bm{m})= &\lambda \cdot \mathcal{L}_{Dice} + (1-\lambda) \cdot \mathcal{L}_{SoftCE},
\end{aligned}
\end{equation}
where $\lambda$ is the weight factor. In our experiments, $\lambda$ is set to 0.5. The term $\mathcal{L}_{Dice}$ measures the overlap between the predicted segmentation and the ground truth, denoted as
\begin{equation}
\begin{aligned}
\mathcal{L}_{Dice} = 1-\frac{1}{N}\sum_{c=1}^{N}\frac{2\sum_{h,w}m_{h,w,c}\hat{p}_{h,w,c}}{\sum_{h,w}m_{h,w,c}+\sum_{h,w}\hat{p}_{h,w,c}},
\end{aligned}
\end{equation}
where $N$ is the number of segmentation classes,  $\hat{p}_{h,w,c}=\frac{e^{{\hat{m}_{h,w,c}}}}{\sum_{k=1}^{N}e^{\hat{m}_{h,w,k}}}$ is the probability distribution obtained by applying the softmax normalization to $\bm{\hat{m}}$ and $m_{h,w,c}$ represents whether each pixel $(h,w)$ in the image belongs to class $c$. The term $\mathcal{L}_{SoftCE}$ is a smoothed version of the cross-entropy loss, a standard objective function for multi-class pixel-level classification tasks, which computes the cross-entropy between the predicted probabilities and the ground truth labels. It can be denoted as 
\begin{equation}
\begin{aligned}
\mathcal{L}_{SoftCE} = -\frac{1}{H \cdot W} \sum_{h,w} \sum_{c=1}^{N}\left[(1-\varepsilon)m_{h,w,c}+\frac{\varepsilon}{N}\right]\text{log}{\hat{p}_{h,w,c}},
\end{aligned}
\end{equation}
where $H$ and $W$ represent the height and width of images, and $\varepsilon$ is smoothed factor, which is set to 0.1 in our task.

\section{Experiment Results}
In this section, we validate the performance of the proposed ProMSC-MIS. Section IV-A outlines the experimental settings. Sections IV-B and IV-C evaluate the performance of ProMSC-MIS against two benchmarking pipelines. In Section IV-D, the visualization results for image segmentation are presented. Finally, Section IV-E compares the model parameters and computational complexity of ProMSC-MIS
with the benchmarks.

\subsection{Experiment Settings}
\subsubsection{Datasets} ProMSC-MIS is trained and tested on MFNet \cite{IEEEhowto:ha2017mfnet}, a real-world public dataset of autonomous driving scenes comprising 1,569 pairs of precisely aligned RGB and thermal (RGB-T) images—820 captured during the daytime and 749 at night. The dataset is divided into a training set, a testing set and a validation set according to the ratio of 2:1:1. 
The images have a resolution of 480 × 640 pixels. The dataset includes eight hand-labeled object classes and one unlabeled background class.

\subsubsection{Training Details and Hyperparameters} We adopt the Adam optimizer for training with a batch size of 4. A step-wise learning rate schedule is used, where the initial learning rate is set to $1 \times 10^{-4}$. The learning rate decays by a factor of 0.9 every 20 epochs. The total number of training epochs is set to 300. All experiments are performed on an Intel Xeon Platinum 8383C CPU, and a 48 GB Nvidia L40 graphics card with Pytorch with CUDA 12.4. 

\subsubsection{Channel Settings} As mentioned before, we adopt the SSCC design where the bit sequence $\bm{b}$ goes through channel encoding, wireless transmission, and channel decoding, resulting in $\bm{\hat{b}}$. To focus on the impact of channel bandwidth limitation, we assume capacity-achieving channel codes are employed for all the considered schemes so that the bit transmission is error-free.

\subsubsection{Benchmarks} 
We consider two benchmarking pipelines: traditional benchmarks and existing deep learning based semantic communication (DeepSC) benchmarks. The former is mainly to demonstrate the advantages of semantic transmission principle of ProMSC-MIS against conventional image transmission, while the latter mainly serves as ablation studies to validate the pre-training and semantic fusion strategies in ProMSC-MIS. 

The traditional benchmarks incorporate traditional image codecs (JPEG2000 and BPG) and existing segmentation models, denoted as JPEG2000/BPG-Seg. Specifically,  images are compressed by JPEG2000 or BPG, transmitted over a wireless channel and reconstructed at the receiver. Image segmentation is then performed on the reconstructed images using deep learning models specifically designed for MFNet dataset, as detailed below.
\begin{itemize}
    \item\textit{MFNet \cite{IEEEhowto:ha2017mfnet} :} This is the first method proposed for multi-spectral image segmentation on the MFNet dataset. The network employs a lightweight architecture with a small number of convolutional layers, resulting in a low parameter count and low computational complexity.
    \item \textit{RTFNet \cite{IEEEhowto:sun2019rtfnet} :} This method utilizes ResNet-152 as its backbone and introduces a multi-scale feature fusion strategy to enhance performance. Compared to MFNet, RTFNet achieves higher accuracy at the cost of increased model parameters and computational complexity.
    \item \textit{FEANet \cite{IEEEhowto:deng2021feanet} :} This method employs an attention mechanism to better capture cross-modal features building upon RTFNet, further improving task performance. It maintains a model complexity and parameter count comparable to that of RTFNet.
\end{itemize}

Note that JPEG2000 and BPG do not provide encoding schemes for thermal images, so these images are compressed as grayscale.

The DeepSC benchmarks comprise two variants: multimodal DeepSC and unimodal DeepSC.
\begin{itemize}
    \item \textit{Multimodal DeepSC:} This benchmark follows the standard multimodal semantic communication pipeline, including unimodal semantic encoding, semantic fusion, wireless transmission, and semantic decoding. In multimodal DeepSC, we adopt the architecture of ProMSC-MIS, reusing all its modules while omitting the pre-training stage. In other words, multimodal DeepSC can be regarded as a baseline to evaluate the impact of the proposed pre-training strategy in ProMSC-MIS.
    \item \textit{Unimodal DeepSC:} This benchmark adopts similar procedure to Multimodal DeepSC, but only processes single modality, therefore is called Unimodal DeepSC. To ensure a fair comparison, we deepen the NN layers in the unimodal semantic encoder so that this scheme maintains similar computation complexity with ProMSC-MIS.
    This benchmark is denoted as RGB-DeepSC or Thermal-DeepSC, depending on the input modality.
\end{itemize}

\begin{figure*}[t]
  \centering
  \begin{subfigure}{0.49\linewidth}
    \centering
    \includegraphics[width=\linewidth]{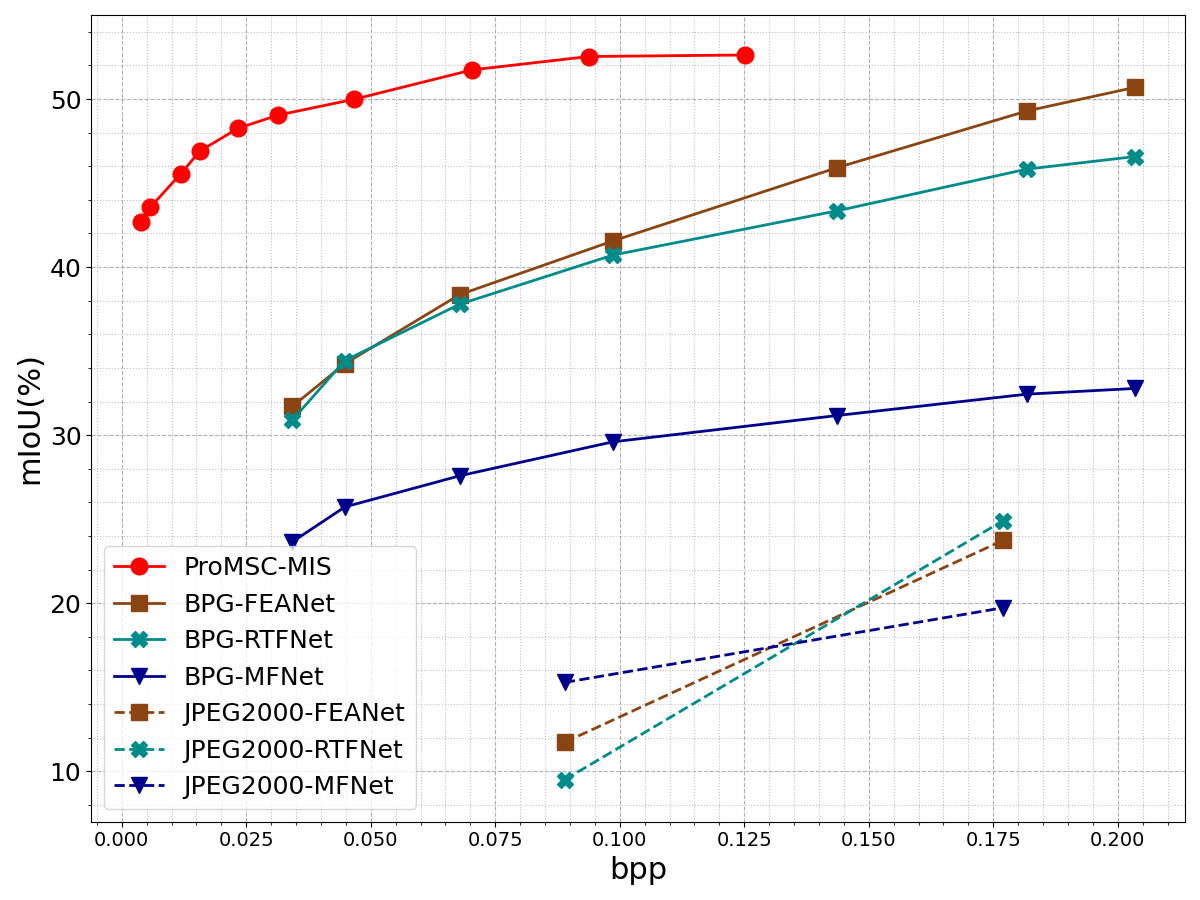}
    \caption{mIoU metric}
    \label{fig:miouvsbpp}
  \end{subfigure}
  \hfill
  \begin{subfigure}{0.49\linewidth}
    \centering
    \includegraphics[width=\linewidth]{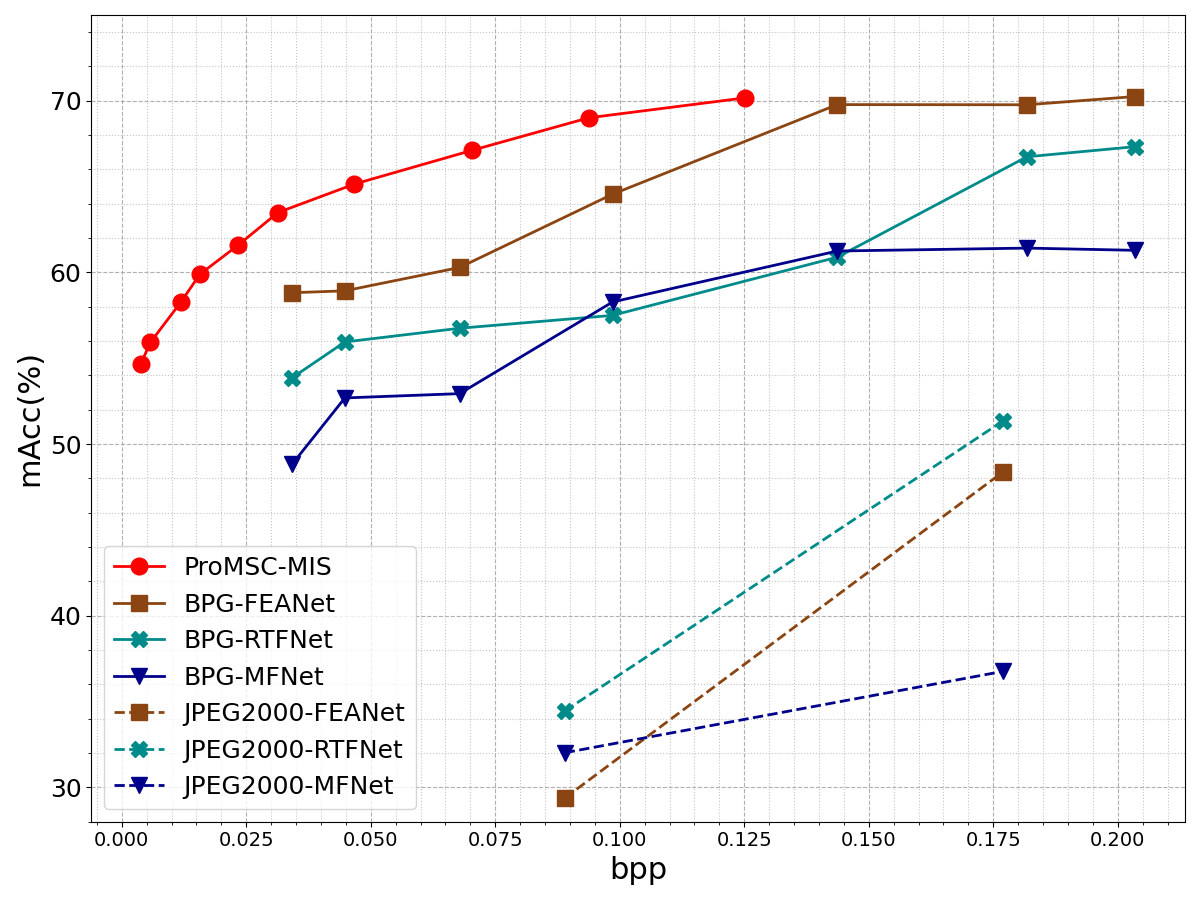}
    \caption{mAcc metric}
    \label{fig:maccvsbpp}
  \end{subfigure}
  \caption{Performance comparison between ProMSC-MIS and JPEG2000/BPG-Seg methods under different bpp levels.}
  \label{fig:metrics_vs_bpp}
\end{figure*}

\subsubsection{Performance Metrics} We adopt mean Intersection over Union (mIoU) and mean Pixel Accuracy (mAcc) as the evaluation metrics. The mIoU quantifies the overlap between $\bm{m}$ and $\bm{\hat{m}}$ while the mAcc measures the classification accuracy for each class individually. 
The two metrics are defined as:
\begin{equation}
    mIoU = \frac{1}{N} \sum_{i=1}^{N} \frac{TP_i}{TP_i + FP_i + FN_i},
\end{equation}
\begin{equation}
    mAcc = \frac{1}{N} \sum_{i=1}^{N} \frac{TP_i}{TP_i + FN_i},
\end{equation}
where $TP_i$ represents the number of pixels correctly predicted as class $i$, $FP_i$ represents the number of pixels incorrectly predicted as class $i$ and $FN_i$ is the number of pixels that belong to class $i$ but are misclassified as other classes.  
The channel-source compression rate is measured in channel bits per pixel (bpp), defined as $L_b / (H \times W)$.


\subsection{Performance Comparison with JPEG2000/BPG-Seg} Fig. \ref{fig:metrics_vs_bpp} shows the performance comparison between ProMSC-MIS and the traditional JPEG2000/BPG-Seg methods across various channel-source compression levels. 

As depicted in Fig. \ref{fig:metrics_vs_bpp}(a), ProMSC-MIS significantly outperforms traditional benchmarks in efficiency, feasibility under extreme conditions and robustness to varying channel bandwidths. First, in terms of efficiency, our method achieves a target performance level with significantly less bandwidth. Notably, to achive an mIoU of $42\%$, ProMSC-MIS requires merely $\frac{1}{20}$ of the bandwidth needed by BPG-FEANet. At a fixed bpp, this efficiency translates into superior task performance. For instance, at a bpp of 0.09375, our method achieves a 3- to 4-fold improvement compared to JPEG2000-based benchmarks (i.e., JPEG2000-MFNet, -RTFNet, and -FEANet), and a 45$\%$ improvement over the more competitive BPG-FEANet. Second, ProMSC-MIS maintains feasibility in extreme low-bitrate regimes (e.g., 0.0039–0.015625 bpp). In this range, conventional codecs like BPG and JPEG2000 fail to reconstruct usable images, making any downstream task impossible, whereas our method still works. Finally, our method exhibits superior robustness to varying channel bandwidths. Traditional benchmarks rely on accurate pixel-level reconstruction, whose quality degrades significantly with bandwidth fluctuations, ultimately impairing segmentation performance. In contrast, ProMSC-MIS benefits from end-to-end optimization that directly obtains segmentation results from the learned representation. This makes it more resilient to channel variations.
Together, these results underscore  key advantages of ProMSC-MIS in highly bandwidth-constrained communication scenarios.


Regarding mAcc shown in Fig.~\ref{fig:metrics_vs_bpp}(b), ProMSC-MIS also shows significant improvements. At 0.09375 bpp, it provides a 2-fold mAcc enhancement over the JPEG2000-based methods. Against BPG-based benchmarks such as BPG-RTFNet and BPG-MFNet, ProMSC-MIS yields an approximate $50\%$ increase in mAcc within the 0.0375 to 0.125 bpp range. Furthermore, when compared with BPG-FEANet, our method delivers similar mAcc performance ($60\%$ to $65\%$) while consuming half the bandwidth. However, it is observed that for higher mAcc values, the performance improvement margin of ProMSC-MIS tends to narrow. This suggests that the its architecture may approach a performance ceiling for mAcc at higher bpp rates, where additional bits bring only slight gains. Despite this observation, even in this regime, ProMSC-MIS generally retains a bandwidth-saving advantage over BPG-FEANet for achieving comparable mAcc levels. 

\begin{figure*}[t]
  \centering
  \begin{subfigure}{0.49\linewidth}
    \centering
    \includegraphics[width=\linewidth]{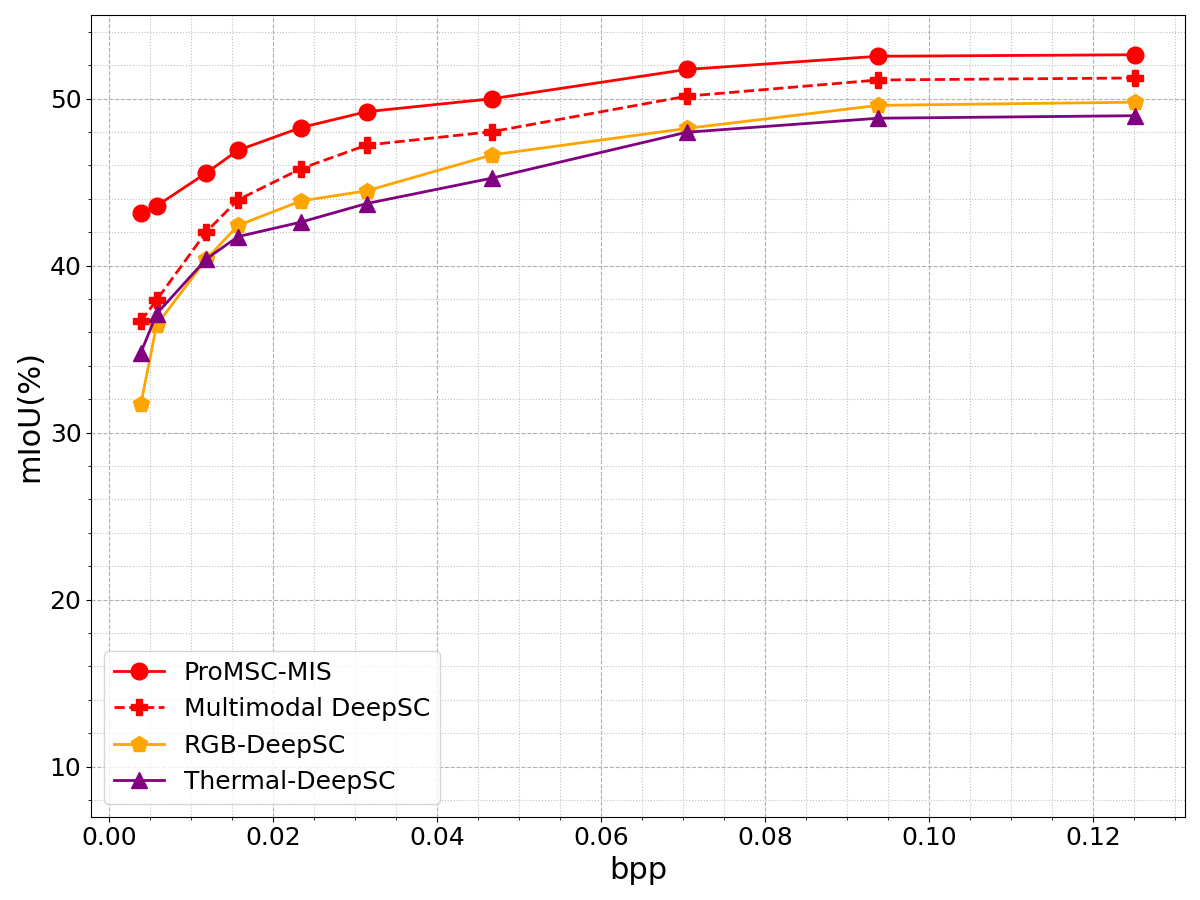}
    \caption{mIoU metric}
    \label{fig:singlemodal_miouvsbpp}
  \end{subfigure}
  \hfill
  \begin{subfigure}{0.49\linewidth}
    \centering
    \includegraphics[width=\linewidth]{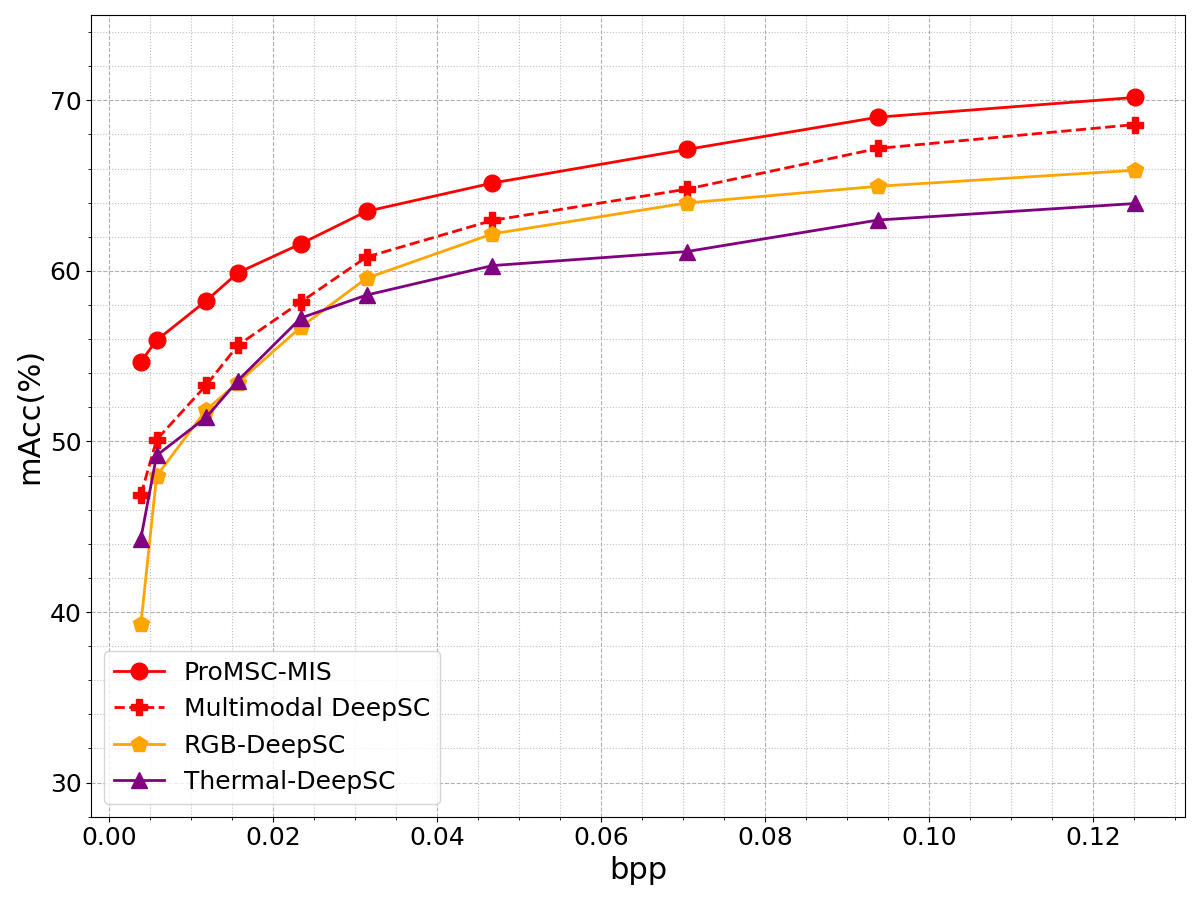}
    \caption{mAcc metric}
    \label{fig:singlemodal_maccvsbpp}
  \end{subfigure}

  \caption{Performance comparison between ProMSC-MIS and DeepSC methods under different bpp levels.}
  \label{fig:ablation}
\end{figure*}

\subsection{Performance Comparison with DeepSC}
Fig. \ref{fig:ablation} shows the performance comparison between ProMSC-MIS and the DeepSC benchmarks.
The experimental results demonstrate that ProMSC-MIS consistently outperforms the multimodal DeepSC. This performance gain is most pronounced at low bpp levels, and  gradually diminishes as the bpp increases. 
This is because at low bpp levels, multimodal features extracted with pre-training contain less redundancy and richer semantic information compared to those extracted without pre-training. With higher bpp levels, the sufficient bandwidth allows models to capture adequate semantic information even without pre-training, reducing the relative advantage. These findings validate the effectiveness of the proposed pre-training approach, particularly in bandwidth limited scenarios.

Moreover, by comparing ProMSC-MIS with the unimodal DeepSC, we can quantify the contribution of each modality to overall task performance. At low bpp levels, the performance gap between ProMSC-MIS and RGB-DeepSC is larger than that between ProMSC-MIS and Thermal-DeepSC. As the bpp increases, however, this gap for RGB-DeepSC gradually narrows and eventually becomes smaller than that for Thermal-DeepSC.
This trend can be attributed to the nature of the data: RGB images contain richer details, which are inherently more challenging to capture effectively under limited bit budgets. As the bpp increases, the RGB model benefits from higher representation capacity, enabling it to better leverage this detailed information and achieve improved performance. In addition, the results also indicate that the method with multimodal data consistently outperforms both RGB-DeepSC and Thermal-DeepSC models at all evaluated bpp levels, regardless of whether pre-training is applied. This strongly demonstrates the potential of multimodal data to improve task performance.


These findings provide practical guidance for the design of both unimodal feature extraction networks and semantic fusion strategies. For instance, the system should allocate more resources (e.g., network complexity) to the RGB semantic encoder and pay more attention to RGB semantic features during semantic fusion, especially when sufficient bandwidth is available.

\subsection{Visualization Results}
\begin{figure*}[t]
  \centering
  \includegraphics[width=\linewidth]{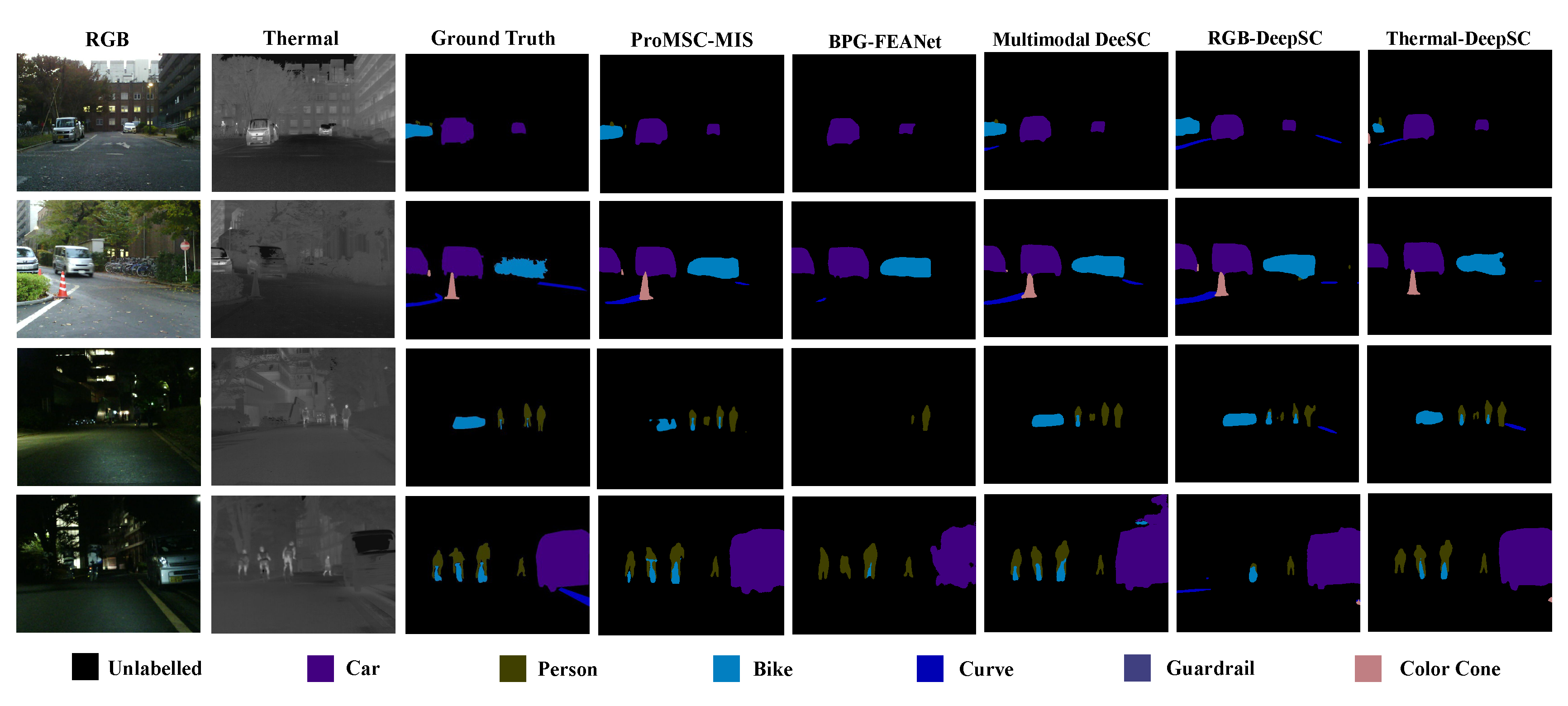}
  \caption{Visualization results of ProMSC-MIS (wtih bpp of 0.0703) and the benchmarks (the traditional benchmarks with bpp of 0.0679 and the DeepSC benchmarks with bpp of 0.0703). Two daytime images and two nighttime images are selected from the test set respectively,  containing a total of seven categories: the unlabelled background class, car, person, bike, curve, guardrail and color cone.}
  \label{fig:semantic_vis}
\end{figure*}

Fig. \ref{fig:semantic_vis} shows the visualization of the proposed ProMSC-MIS and benchmarks at similar bpp levels.
The visualized samples consist of two daytime and two nighttime RGB-T image pairs from the test set and cover six prevalent semantic categories and a background class.  For the traditional benchmarks, we display only the best-performing method which is BPG-FEANet.

Compared to the traditional BPG-FEANet method, ProMSC-MIS consistently generates results with sharper object contours and more comprehensive semantic categories. This advantage is evident in the daytime scene (the second row), where ProMSC-MIS accurately identifies and segments the ``color cone" region---an instance entirely missed by BPG-FEANet. The performance gap between the two methods becomes even more pronounced in the more challenging nighttime scenarios. For instance, in the fourth row, ProMSC-MIS successfully detects all ``bike" objects, whereas BPG-FEANet fails to fully recognize them. This detail is critical for correct scene interpretation: our method enables the recognition of ``a person riding a bike", while the benchmark might misinterpret the scene as ``a person walking".

The advantage of ProMSC-MIS also extends to its comparison with the DeepSC benchmarks. When compared against multimodal DeepSC, our method's primary advantage lies in its significantly more accurate object contours. As shown in the fourth row, the ``car" segmentation produced by ProMSC-MIS closely aligns with the ground truth, whereas the corresponding region from multimodal DeepSC suffers from noticeable distortion. Furthermore, a comparison with the unimodal DeepSC highlights the complementary nature of RGB and thermal data. In well-lit daytime scenes, the RGB modality is more informative. For example (the second row), RGB-DeepSC captures a more complete set of classes, while Thermal-DeepSC fails to identify the ``curve", which is also nearly invisible in the original thermal image. Conversely, in nighttime scenes, the thermal modality excels. In the fourth row, Thermal-DeepSC leverages thermal signatures to correctly segment the ``person riding a bike" scene, a detail that RGB-DeepSC completely fails to distinguish due to poor lighting. 

The qualitative improvements in visual fidelity shown here are consistent with the superior quantitative performance metrics presented in Fig.~\ref{fig:metrics_vs_bpp}. These visualizations clearly demonstrate the complementary nature of RGB and thermal modalities: the RGB images provide rich details in daytime scenes, while thermal images preserve object information in low-light conditions.

\newcolumntype{Y}{>{\centering\arraybackslash}m{\dimexpr\linewidth/4-2\tabcolsep\relax}} 
\begin{table}[t]
\small
\centering
\caption{Comparison of Parameters, FLOPs and Inference Latency}
\label{tab:comparison}
\setlength{\tabcolsep}{3pt} 
\renewcommand{\arraystretch}{1.5} 
\begin{tabularx}{\linewidth}{ Y | Y | Y | Y } 
\Xhline{0.8pt}
Model & Params (M) & FLOPs (G) & Inference Latency (ms)\\
\Xhline{0.8pt} 
ProMSC-MIS (Average) & 186.99 & 212.34 & 46.95 \\
\hline
MFNet & 0.74 & 8.42 & 4.60 \\
\hline
RTFNet & 254.51 & 337.46 & 51.87 \\
\hline
FEANet & 255.21 & 337.47 & 65.89 \\
\Xhline{0.8pt}
\end{tabularx}
\end{table}

\subsection{Model Parameters and Computational Complexity}
Table \ref{tab:comparison} shows comparison of parameter counts, floating point operations (FLOPs) and inference latency per RGB-T image pair between ProMSC-MIS and the traditional benchmarks.
The complexity of our method varies with bpp levels, with parameters ranging from 165.27M to 216.3M, FLOPs from 168.21G to 261.86G and inference latency from 37.38ms to 53.18ms.
On average, ProMSC-MIS achieves lower computational cost and faster inference than RTFNet and FEANet. Even though MFNet has the lowest complexity and smallest parameter count, its performance is much inferior to others as indicated in Fig. \ref{fig:metrics_vs_bpp}. This relatively low complexity mainly stems from its simpler CNN architecture with fewer layers. 

It is worth noting that the computational complexity of the benchmarks listed in Table \ref{tab:comparison} has not yet included the cost of the traditional source coding (JPEG2000 and BPG), while the complexity of our ProMSC-MIS accounts for the entire pipeline.  As such, from the overall perspective, it is clear that ProMSC-MIS achieves significantly better task performance while maintaining low computational complexity and storage overhead.

\section{Conclusion}
In this paper, we propose ProMSC-MIS, a prompt-based multimodal semantic communication system for multi-spectral image segmentation. Drawing inspiration from prompt and contrastive learning, we introduce a novel pre-training strategy to guide unimodal encoders in learning diverse semantic features. Furthermore, we employ cross-attention mechanisms and SE networks to facilitate efficient semantic fusion. Experimental results demonstrate the efficiency and effectiveness of ProMSC-MIS. First, compared with the JPEG2000/BPG-Seg methods, ProMSC-MIS reduces the required bandwidth 50\%--70\% at the same segmentation performance, while also decreasing the storage overhead and computational complexity by 26\% and 37\%, respectively. Second, compared with DeepSC methods, ProMSC-MIS consistently outperforms both multimodal DeepSC without pre-training and unimodal DeepSC, validating the effectiveness of the proposed pre-training and semantic fusion strategies. In addition, a comparison with unimodal DeepSC reveals the distinct contribution of each modality to task performance at different compression levels, offering practical insights for uniodal encoders and fusion strategy design.


%

\ifCLASSOPTIONcaptionsoff
  \newpage
\fi



\bibliographystyle{IEEEtran}
\bibliography{IEEEabrv,paper}
\end{document}